# OPTICAL CONTROL OF MASS EJECTION FROM FERROELECTRIC LIQUID DROPLETS: A POSSIBLE TOOL FOR THE ACTUATION OF COMPLEX FLUIDS

Stefano Marni[1], Raouf Barboza[1], Annamaria Zaltron[2] and Liana Lucchetti[1]*

[1]Dipartimento SIMAU, Università Politecnica delle Marche, via Brecce Bianche, 60131 Ancona, Italy.

[2] Dipartimento di Fisica e Astronomia G. Galilei, Università di Padova, via Marzolo 8, Padova, Italy

E-mail: l.lucchetti@univpm.it

## Abstract

We report on the optical control of the recently observed electromechanical instability of ferroelectric liquid droplets exposed to the photovoltaic field of a lithium niobate ferroelectric crystal substrate. The ferroelectric liquid is a nematic liquid crystal in which almost complete polar ordering of the molecular dipoles generates an internal macroscopic polarization locally collinear to the mean molecular long axis. Upon entering the ferroelectric phase, droplets irradiated by unfocused beam undergo an electromechanical instability and disintegrate by the explosive emission of fluid jets. We show here that the regions of jets emission can be controlled by focusing the light beam in areas close to the droplet's edge. Once emitted, the fluid jets can be walked by moving the beam up to millimeter distance from the mother droplet. Reverting the lithium niobate substrate, jets become thinner and show the tendency of being repelled by the beam instead of being attracted, thus offering an additional tool for their optical manipulation. These observations may pave the way to intriguing applications of ferroelectric nematic fluids related to manipulation, actuation, and control of soft, flexible materials.

## Introduction

The recent discovery of the ferroelectric nematic phase, $N_F$, [1], opened a new chapter not only for the liquid crystal community, but in the whole condensed-matter physics. Beside adding a new, very peculiar, member to the group of ferroelectric materials, the new phase offers a broad range of physical effects to explore, ranging from the behavior of topological defects to surface anchoring [2], response to low frequency electric fields [3] and light, interplay of bound and free electric charges, viscoelastic properties, field-controlled hydrodynamics [3-6], field-order coupling in both the $N_F$ and the pre-transitional regions, behavior in confined geometry [7], just to cite a few examples. In this scenario, we recently performed experiments devoted to characterizing the behavior of sessile $N_F$ droplets on ferroelectric solid substrates [4] and found that the combination of fluidity and polarity gives rise to an electromechanical instability induced by the coupling of the LC polarization with the polarization of the solid substrate. This latter can be induced either pyroelectrically by temperature variation [4] or photovoltaically through light irradiation [8].

In both cases jets ejection starts from the droplet rim where topological defects more easily form and polarization charge is easily accumulated [4,8]. This suggests the possibility of controlling the phenomenon by irradiating specific parts of the ferroelectric nematic droplet, which is the focus of this work. Indeed, the advantages of using light are related to the possibility of focusing the incident beam to small regions of the substrate thus limiting the extent of the charged regions, and of easily controlling its position with respect to the droplet. Moreover, the illuminated area can be quickly reconfigured in different patterns and geometries, possibly allowing all the basic protocols of fluid manipulation. While in conditions of isotropic irradiation of the whole droplet, the observed instability consists of the emission of fluid jets from randomly distributed positions along the droplet rim [4,8], a beam focused close to the droplet edge is expected to induce the ejection of a single fluid jet from that specific position. Our results confirm this hypothesis and show that the ejected jets can be optically manipulated in different ways. Specifically, the fluid ferroelectric jets can be walked over long distances by changing the position of the beam on the substrate, before breaking up, partly going back to the mother droplet, and partly forming small secondary drops. Moreover, by depositing the ferroelectric droplets on the other side of the substrate, jets become thinner and show the contrary tendency to be repelled away from the illuminated area, which can also be used for controlling their motion and for jet splitting and redirection.

Worthy of note the behavior of liquid crystalline materials in combination with photoactivable substrates has also been studied in the case of the conventional nematic phase in several different configurations [9-12]. The effect of the substrate charging on the LC average molecular orientation has been demonstrated, however no effect comparable to those observed with ferroelectric nematic LCs was reported.

**Materials and methods**

The ferroelectric liquid crystal used in this work is 4-[(4-nitrophenoxy)carbonyl]phenyl2,4-dimethoxybenzoate (RM734). It was synthesized as described in [1] and its structure and phase diagram have already been reported [1, 2, 4]. In this compound the ferroelectric nematic phase appears through a second order phase transition upon cooling from the conventional higher temperature nematic (N) phase and exists in the range 133°C - 80°C [2,4]. The spontaneous polarization **P** of RM734 is either parallel or antiparallel to the molecular director **n,** defining the average orientation of the molecular axis**,** and exceeds 6 $\mu C/cm^2$ at the lowest T in the $N_F$ phase [1].

As ferroelectric solid substrates we used 900 µm thick z-cut lithium niobate (LN) crystals. Experiments were performed on iron-doped substrates containing 0.1% mol. of iron with a reduction factor R, defined as the ratio $Fe^{2+}/Fe^{3+}$, of 0.02. The bulk spontaneous polarization $P_{LN}$ of LN crystals along the [0001] z-axis is of the order of 70 $\mu C/cm^2$ and does not depend significantly on T in the explored range since its Curie temperature is much higher (≈ 1140°C). The huge bulk polarization of LN does not however translate in a huge surface charge density because of very efficient compensation mechanisms at the z-cut surfaces, lowering the equilibrium surface charge to only about $10^{-2}$ $\mu C/cm^2$ [13]. When the crystal is exposed to light with a wavelength in the iron absorption spectrum, the surface charge of LN significantly increases because of the photovoltaic effect [14], consisting in the appearance of a photo-induced current according to the scheme $Fe^{2+} + h\nu \rightarrow Fe^{3+} + e^-$. The subsequent charge distribution that takes place inside the crystal gives rise to an internal electric field, the photovoltaic field, with saturation values that can reach $10^7$ V/m, depending on the dopant concentration and on R [14]. In the case of LN crystals containing 0.1% mol. of iron and having R<<1, as those used here, the photovoltaic field was indeed calculated to be of the order of $10^7$ V/m [15]. Taking into account the proper dielectric permittivity of LN, we can thus expect an induced surface charge density of the order of 1 $\mu C/cm^2$. LN crystals were used as bare substrates with no coating applied. Experiments have been performed on both the crystal sides.

The RM734 droplets were obtained as described in the SI and have diameter in the range (200 – 400) µm, as measured with a calibration slide. They were deposited on bare LN substrates that were previously slowly heated up to T = 200°C, corresponding to the RM734 isotropic phase. Successively, T was decreased down to a value in the range (130 – 105) °C, where RM734 is in the $N_F$ phase. Noteworthy, the cooling rate was kept slow enough to avoid the droplets electromechanical instability observed in [4] that was triggered by the pyroelectric charging of LN surfaces and required a proper cooling speed.

The light used to induce the photovoltaic effect in LN crystals is a gaussian beam from a frequency doubled Nd:YAG laser ($\lambda$ = 532 nm) focused to a waist w = 35 µm, with intensity ranging from 0.5 x $10^2$ W/$cm^2$ to 2 x $10^2$ W/$cm^2$. LN substrates were irradiated from below in different positions close to the droplet edges, holding the temperature fixed. Polarized optical microscope (POM) observations during light irradiation were carried out and videos of the droplets behavior were recorded with a rate of 25 frames per second.

**Results**

As we recently reported [8], when the LN substrate is irradiated in correspondence of the droplet position with an unfocused gaussian beam slightly larger than the droplet itself, an electromechanical instability very similar to the one reported in [4] is observed. The instability indeed consists in the sudden emission of fluid jets that are mostly interfacial and spread out on the substrate surface exhibiting fractal branching [16,17], as shown in Fig. 1. Jets are emitted from randomly distributed regions along the droplet rim, reflecting the uniform illumination of the droplet. As discussed in [8], we understand this phenomenon in analogy to the interpretation proposed in [4], i.e., as due to the fringing field generated by the photovoltaic charging of LN substrates, which polarizes the $N_F$ droplets. This takes place through a small reorientation of the fluid polarization **P** by an angle such to deposit polarization charge on the droplet top and bottom surfaces that cancel the internal field [4]. The amount of required surface charge is much smaller, by orders of magnitude, than the spontaneous bulk polarization **P** of the RM734 $N_F$ phase, indicating that very small reorientations of **P** are required to screen the internal fringing field and that such screening takes place right at the transition to the $N_F$ phase. In this process, the fringing field in the surrounding space is modified by the arising charge distribution at the droplet/air interface and acquires a planar radial component because of the droplets dome shape. Such radial component appears to drive the motion of the jets at the instability. In all our observations, jets start from the rim of the droplet, which appears to indicate that the edges of the droplets are likely location where topological defects more easily form and polarization charge accumulates, thus acting as trigger points for jet formation [4]. Indeed, in the process of droplet polarization, opposite directions of **P** might nucleate and converge in specific locations thus giving rise to domain walls. A domain wall of surface S accumulates a charge q=2PS. The instability occurs when the Coulomb force $F_C = k_C q Q / R^2$ ($Q=\sigma A$ is the droplet charge and $A$ its surface) becomes larger that the force $F_S$ arising from the surface tension and opposing the formation of a cusp whose vertex is the localized charge pulling away from the droplet, which will be of order of $F_S \sim \gamma \sqrt{S}$ [4].

Since the photovoltaic field in LN crystals is mostly confined within the irradiated area [14], we expect that uniform illumination of the droplet gives rise to uniform charging, thus mimicking the experimental conditions of ref. [4].

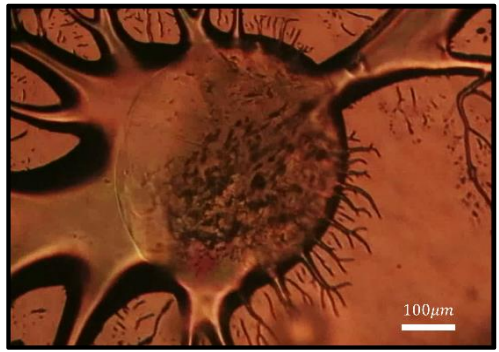

**Figure 1**: RM734 sessile $N_F$ droplet on LN substrate exposed to uniform green light illumination. The beam diameter is slightly larger than that of the droplet. Shape instability with the emission of interfacial fluid jets from randomly distributed regions along the droplet rim, is clearly visible. I = 20 W/cm$^2$, droplet diameter 350 µm, T = 110°C.

This confinement of the photovoltaic charging also limits the extension of the corresponding fringing field. On the other hand, the size and position of the illuminated area can be varied almost at will, as well as the intensity of the light beam. This combination of circumstances, summed to the demonstrated existence of a coupling between the $N_F$ droplet polarization and the photovoltaic fringing field [8], suggests that the droplet instability can be controlled quite precisely in terms of position, by using beams smaller than the droplet diameter and positioning them in specific locations along the droplet edge.

To test this possibility, the light beam was focused to a waist of 35 µm and positioned close to the droplet rim, as in Fig. 2a. In these conditions light irradiation causes a sort of localized instability that consists in the emission of a single fluid jet in correspondence of the irradiated area (Fig. 2b). The delay between the beginning of light irradiation and jet ejection ranges from 0.5 to 20 s depending on both light intensity and initial distance of the light spot from the droplet edge. Specifically, the delay decreases with increasing I and decreasing the initial distance, which highlights the role of photovoltaic charging. The time needed for the photovoltaic field – and thus for the fringing field - to reach the maximum value is indeed inversely proportional to the intensity of the incident light [14]. Moreover, the fringing field strength decreases with increasing distance. After light removal, the fluid jet keeps its position for up to 30 s before disrupting by partly going back to the mother droplet and partly forming secondary small drops, as shown in Fig. 2c and d, where the new droplet is indicated by an arrow. The typical branching with several levels of ramification, characteristic of jets expelled by $N_F$ droplets [4,5], are clearly visible. The maximum distance $r_{MAX}$ between droplet edge and light spot center that allows to observe the effect depends on light intensity, as shown in Fig. 2e, and is of the order of hundreds of microns in the range of intensity used here. Surprisingly, it does not seem to depend on the value of the substrate temperature T, as shown in Fig. 2f, where the occurrence of jet ejection has been plotted as a function of the distance $r$ between light spot center and droplet edge for several values of T within the $N_F$ range, and for I = 80 W/cm$^2$. Interestingly, light irradiation of small regions entirely within the droplet area does not lead to jet ejection. This suggests that the effect of LN charging on droplets polarization varies with both the size and the position of the illuminated charged area. Note that the observed behavior is not necessarily in contradiction with the instability observed in Fig.1, produced by a uniform illumination of the whole droplet/LN interface. Figures 2a - d have been extracted from Video S1, available in the SI.

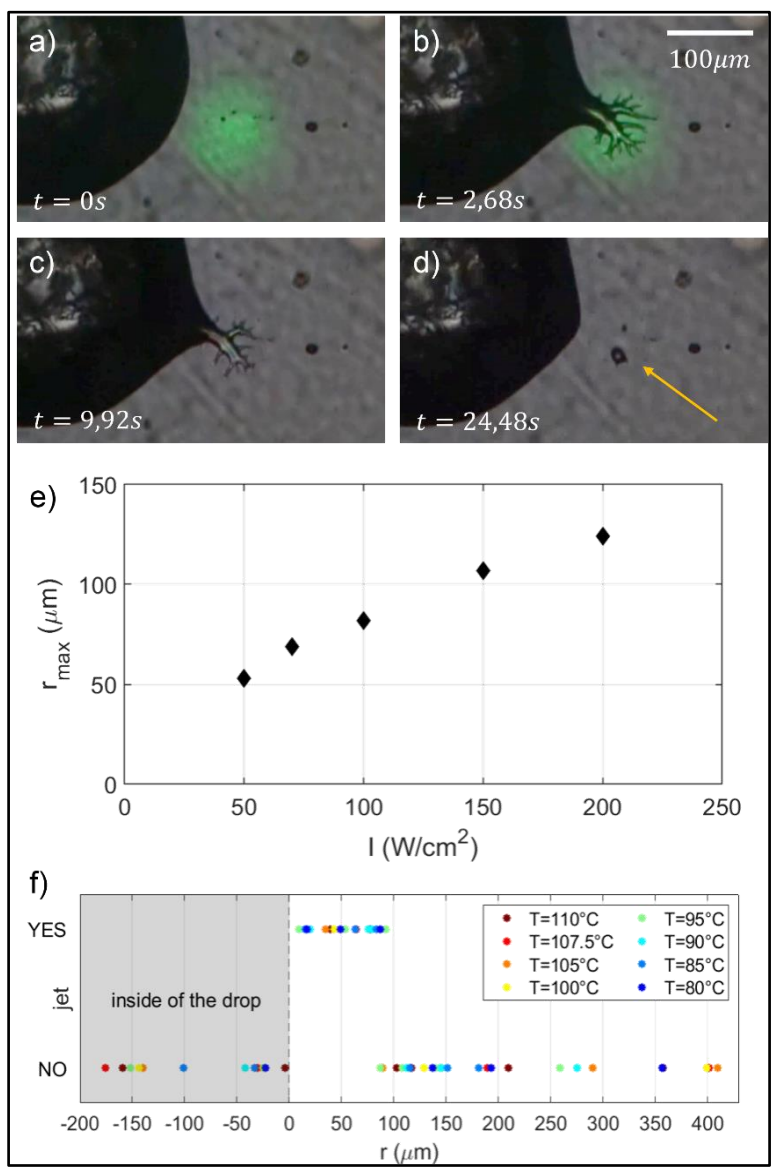

**Figure 2:** Focused irradiation of a LN substrate in a position close to a RM734 ferroelectric sessile droplet. The beam positioned close to the droplet rim (a), produces the emission of a single fluid jet in correspondence of the irradiated area (b). Upon switching off the light, the jet keeps its position for up to 30 s (c) before disrupting by partly going back to the mother droplet and partly forming secondary small drops (d). The yellow arrow indicates the remaining secondary droplet. T = 130 °C, I = 80 W/cm². e) Range of interaction, measured as the maximum distance between the light spot center and the droplet edge, as a function of the light intensity. f) Occurrence of jet ejection as a function of the distance between light spot center and droplet edge, for different values of the substrate temperature and for I = 80 W/cm².

Once the fluid jet has been generated in the chosen position, it can be walked by moving the light beam, that it follows up to distances of thousands of microns, until it breaks up and forms small, secondary droplets. This is shown in Fig. 3 (extracted from Video S2), where the jet's walking occurs through a continuous emission of new streams of fluid from its tip toward the illuminated area. The red light in the figures comes from a low frequency He-Ne laser used for the side imaging of the original droplet, aimed at monitoring the contact angle variations close to the electromechanical instability. The results of this analysis are not relevant for the present study and will be reported in a forthcoming work.

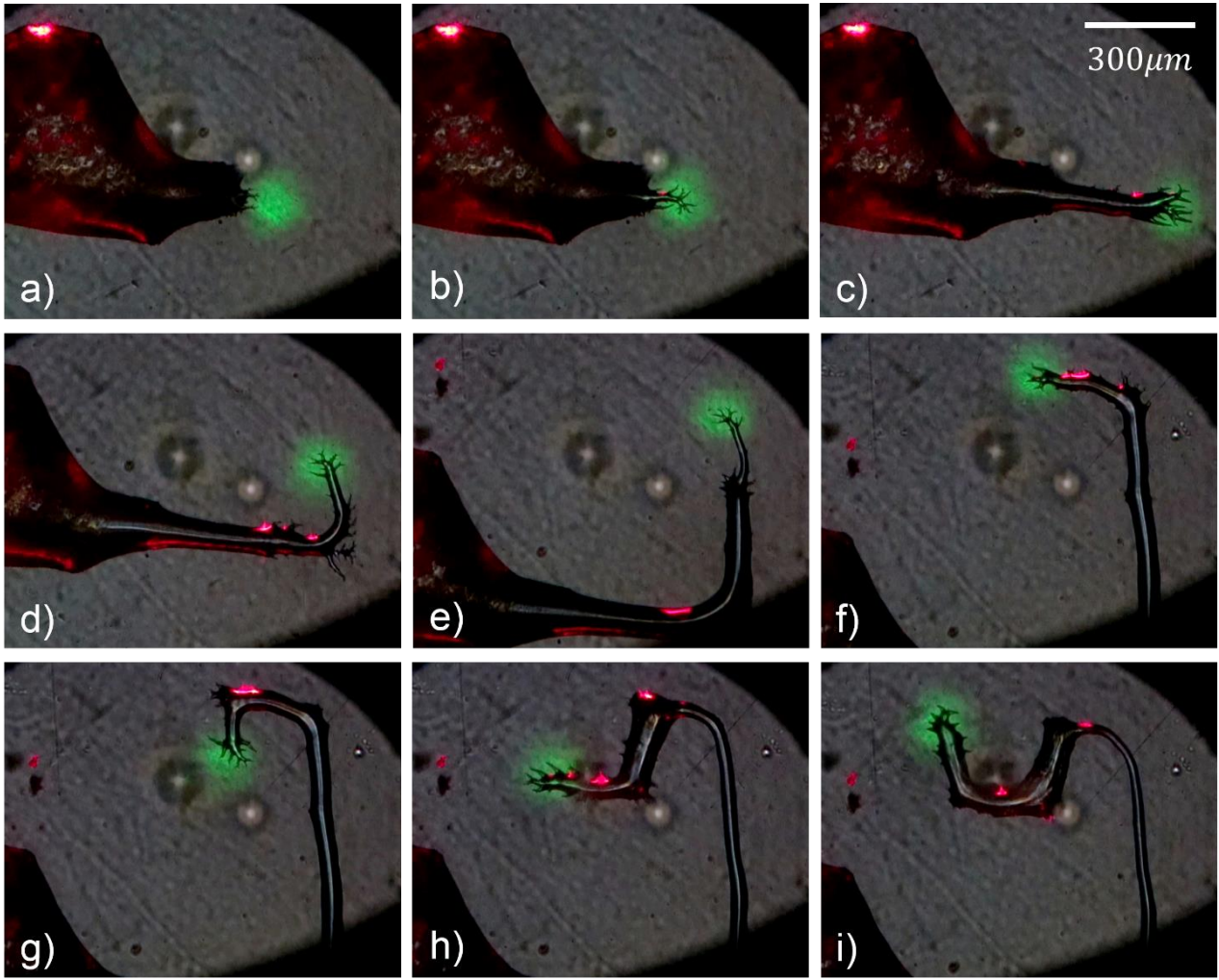

**Figure 3**: Video frames at different instants showing a fluid jet following the motion of the light spot. Jet's walking occurs through a continuous emission of new streams of fluid from the tip toward the illuminated area. The red light seen in the figures comes from a low frequency He-Ne laser used for the side imaging of the original droplet. T = 120°C, I = 100 W/cm$^2$.

Interestingly, while the effect reported in Fig. 1 is independent on the side of the LN crystal contacting the droplet, being thus in agreement with [4] where the sign of the charges of the LN surface that contacts the LC droplet was irrelevant, the localized instability obtained with focused beams behaves in different ways on the two surfaces. Specifically, by depositing RM734 droplets on the opposite LN surface (which we will call the DOWN surface to distinguish it from the UP surface), one obtains thinner jets that exhibit the tendency of being repelled by the illuminated area. This is reported in Fig. 4 where four frames extracted from Video S3 show the light-induced localized instability observed on the DOWN side of the same LN substrate used in Fig. 2 and 3. As described, the fluid jets locally emitted by the N$_F$ droplet are generally thinner and quickly retract when the light is still on, by both going back to the mother droplet and disrupting in small secondary drops.

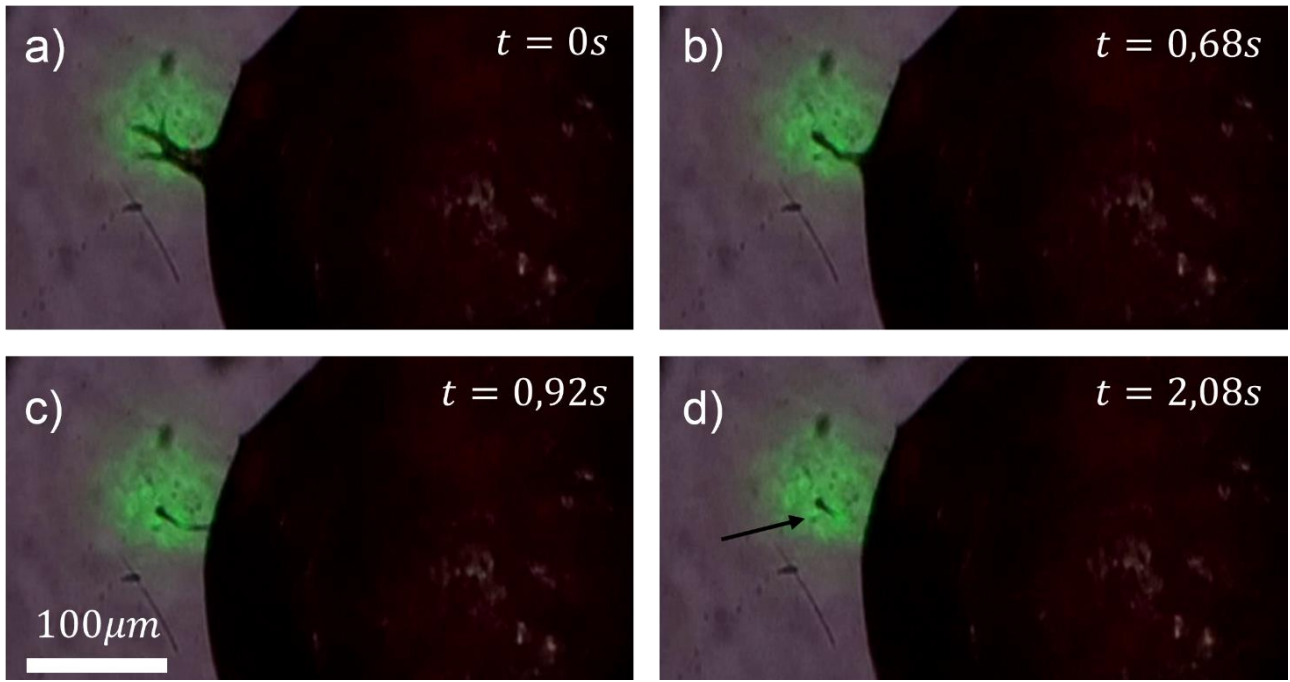

**Figure 4**: Focused irradiation of the same LN substrate used in Fig. 2, in a position close to a RM734 ferroelectric sessile droplet deposited on the opposite (DOWN) side of the crystal. The beam positioned close to the droplet rim produces the emission of a single fluid jet in correspondence of the irradiated area (a). The jet spontaneously retracts back to the mother droplet during light irradiation (b) and (c), leaving behind a small secondary drop indicated by a black arrow (d). T = 120°C, I = 100 W/cm$^2$.

Optical manipulation of fluid jets is observed also in this geometry, although with different features with respect to the jet walking shown in Fig. 3. The tendency of the emitted fluid jets to be repelled by the light spot can indeed be exploited for moving them based on a mechanism which is different from the continuous emission of new branches that move toward the irradiated area. Figure 5 (extracted from video S4) shows an example of this different light induced jet motion. The long jet visible in Fig. 5a has been generated by focused irradiation of a region within the droplet perimeter (which, on the DOWN side, does give rise to jet ejection) and elongates in the direction opposite to the light spot as this latter gets closer to its tip (Fig. 5b). Continuous irradiation of the jet results in a thinning of its section (Fig. 5 c and d) up to the separation from the original droplet (Fig. 5 e). Then, taking advantage of the repulsive force between the jet's tip and the light spot, the "free" fluid jet can be optically redirected as reported in Fig. 5 f, g and h. A close inspection of Video S4 reveals other examples of the same behavior.

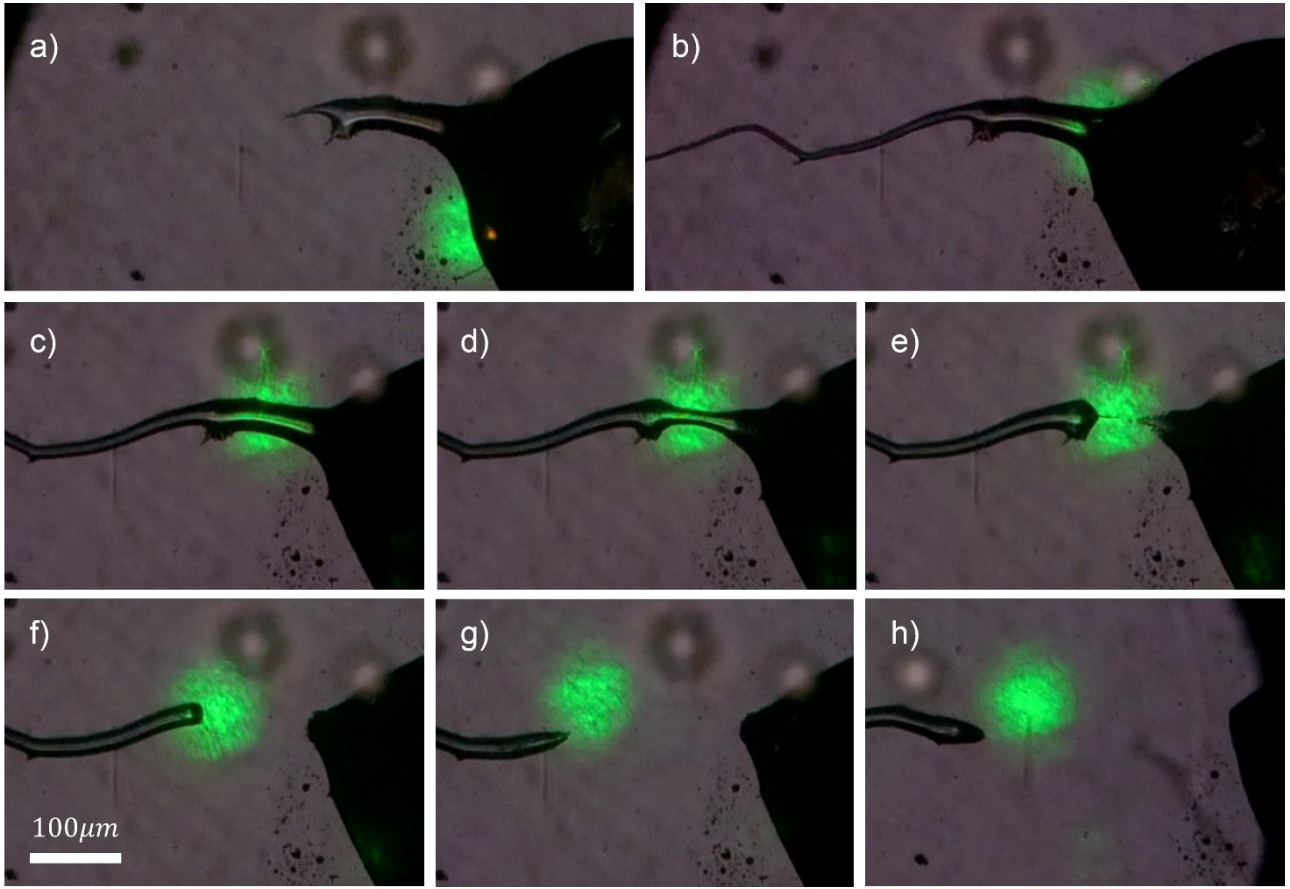

**Figure 5**: Long fluid jet emitted by a sessile RM734 $N_F$ droplet upon focused illumination of an area within the droplet perimeter, on the DOWN LN side (a). By shifting the light spot in correspondence of the droplet rim, the jet elongates (b) while its irradiated section becomes thinner and thinner (c) and (d), up to the separation from the original droplet (e). The motion of the "free" fluid jet can then be optically controlled by moving the light spot close to its tip, taking advantage of the resulting repulsive force (f) –(h). T = 120°C, I = 150 W/cm$^2$.

**Discussion**

The electromechanical instability of sessile droplets on LN substrates, observed both in the presence of pyroelectric [4] and photovoltaic [8] LN charging, is a consequence of the fringing field-mediated coupling between the droplet polarization and the polarization of the underlying substrate. In case of light irradiation, the fringing field is localized in the neighborhood of the illuminated area. Thus, if a light beam smaller than the ferroelectric droplet is positioned close to a specific part of the droplet edge, it mainly affects that specific part of the droplet, which thus becomes the site of polarization-induced charge accumulation that triggers jet ejection. Jets are thus expelled so that their charged tips are as far as possible from the droplet rim, as a strategy to minimize the accumulated electrostatic energy [4]. The result is a motion of the jets toward the illuminated area, as observed in both Fig. 2 (UP side) and Fig. 4 (DOWN side). This behavior is independent on the sign of the surface charges generated on the LN surfaces upon light illumination. However, once the fluid jets are expelled by the mother droplets, they behave differently depending on the specific side of the LN substrate they lie on, being stable for several tens of seconds even in the absence of light on the UP side (Fig. 2) and spontaneously retracting almost immediately on the DOWN side (Fig. 4). Moreover, focused irradiation of the droplets in regions within their perimeter produces jet ejection only on the DOWN side, while does not lead to any evident instability event on the UP side.

The simplest form of the coupling of polarized jet tips to light is through dielectrophoresis [18]. The dielectrophoretic force *F* is proportional to the square of the field gradient through the relation [18]:

$$F = 2\pi R^3 \varepsilon_m Re\left[\frac{\varepsilon_p - \varepsilon_m}{\varepsilon_p + 2\varepsilon_m}\right]\nabla E^2$$

valid in the simple, general case of a spherical particle of radius R and complex dielectric permittivity $\varepsilon_p$, dispersed in a medium of complex dielectric permittivity $\varepsilon_m$. In the case analyzed here, *F* is generated by the non-uniform fringing field present close to the LN surface. We have recently shown that such a dielectrophoretic force increases with increasing light intensity and can put into motion small (tents of microns) RM734 ferroelectric droplets lying on LN substrates coated by a thin layer of fluorolink, which decreases the surface adhesion force [8]. Here, ferroelectric droplets are larger and the LN substrates do not have any coating, so the adhesion forces are strong and oppose the motion of the overall droplet, which indeed is not observed. In this case, the effect of the dielectrophoretic force is on the jets tips, that are attracted or repelled depending on the side of the LN substrate over which they are moving. We remark that jet ejection occurs when the local accumulation of polarization charges gives rise to repulsive forces that become unsustainable by the surface tension [4]. The dielectrophoretic force, which is thus not responsible for jets formation, then drives their motion.

Being *F* proportional to the square of the field gradient, its effect on the jet dynamics depends on the spatial profile of the field amplitude. We thus expect different shapes of the field amplitude on the two LN sides, as also recently observed in [8]. Specifically, the described behaviors suggest a fringing field with a maximum value close to the light spot center on the UP side, which would generate attraction toward the illuminated area, and a fringing field with a maximum value far from the spot center, on the DOWN side, which would instead produce repulsion from the illuminated area. This asymmetric behavior of the LN substrate might be due either to an intrinsic imbalance of surface charging or to a different combination of fields on the two surfaces, both these effects being reported in the literature. Indeed, it has been experimentally observed that the photovoltaic charge accumulation in LN z-cut crystals is not symmetric on the two surfaces, being the -c surface more efficient in generating a strong electrostatic field than the + c surface [19,20]. Moreover, a side-dependent combination of pyroelectric and photovoltaic fields has been shown to produce a different response in the case of conventional nematic liquid crystals confined in microchannels engraved in LN, when light impinged on the +c or on the -c crystal surface [21]. Such a side-dependent combination of photovoltaic field, generated by light irradiation, and pyroelectric field, generated by light absorption, can be at play in our experiments and give rise to different amplitude profiles of the resulting fringing field. Noteworthy, a fringing field with a maximum located far from the light spot center would also give account of the jet "cutting" observed in Fig. 5e and in video S4. Indeed, a repulsive dielectrophoretic force can push the fluid material away from the center of the light spot eventually causing a total loss of mass in correspondence of the illuminated area thus giving rise to the observed break up of one jet into two parts.

Beside producing different field profiles, the supposed side-dependent combination of photovoltaic and pyroelectric fields might also result in a total field having different maximum value [8]. This could affect the role played by the different forces in action: the Coulomb repulsion between domain walls and the rest of the droplet, the surface tension, the dielectrophoretic attraction toward the regions of maximum field and, possibly, the electrophoretic interaction between the charged jets tip and the LN charged area. A different balance among these forces could possibly be at the basis of the observed absence (occurrence) of jets ejection in the case of droplets on the UP (DOWN) LN side, when illuminated within their perimeter. Work is in progress to characterize the side-dependent LN behavior.

**Conclusions**

We demonstrated the possibility of optically controlling the electromechanical instability recently observed in ferroelectric liquid droplets exposed to the polarization field of a ferroelectric solid substrate. The instability leading to the ejection of fluid jets can be induced in controlled regions along the droplet rim and the emitted jets can be optically manipulated by imposing them specific path along the substrate. The optical actuation is realized either through the attraction between the jets tip and the center of the light spot, which results in walking the jet over long distances, or through the repulsion, which gives rise to equally intriguing effects as jets elongation away from the illuminated area or jet splitting. We interpreted the observed effects as a combination of localized electromechanical instability and dielectrophoretic-driven motion of the expelled fluid jets. The former is the result of the Coulomb repulsion between the charge at the droplet/air interface and the charge accumulated at the defects and occurs as this electrostatic force overcomes the surface tension. Both these phenomena are due to the fringing field generated by LN charging that mediates the coupling between the polarizations of the fluid and solid ferroelectric materials.

The range of interaction between the ferroelectric liquid and the illuminated spot can be controlled by acting on light intensity, which gives a versatile degree of freedom for the control of the jet ejection and motion. On the other hand, it is not instead affected by the substrate temperature, which appears a desirable feature in view of possible applications.

Our observations are a consequence of the peculiar combination of fluidity and polar coupling to electric fields that characterizes the recently discovered ferroelectric nematic phase. In this sense, the results obtained contribute an additional piece to the understanding of this new phase of matter and might open the way to interesting applications related to the optical actuation of complex fluids.

**Acknowledgments**

We wish to thank Noel Clark and Matt Glaser for providing the liquid crystal sample and for useful discussions.

**Captions**

**Figure 1:** RM734 sessile N$_F$ droplet on LN substrate exposed to uniform green light illumination. The beam diameter is slightly larger than that of the droplet. Shape instability with the emission of interfacial fluid jets from randomly distributed regions along the droplet rim, is clearly visible. I = 20 W/cm$^2$, droplet diameter 350 µm, T = 110°C.

**Figure 2:** Effect of **LN** focused irradiation close to a RM734 sessile droplet in the $N_F$ phase. T = 130 °C, I = 100 W/cm$^2$. The beam positioned close to the droplet rim (a), produces the emission of a single fluid jet in correspondence of the irradiated area (b). Upon switching off the light, the jet keeps its position for up to 30 s (c) before disrupting by partly going back to the mother droplet and partly forming secondary small drops (d). The blue arrow indicates the remaining secondary droplet. e) and f) Range of interaction, measured as the maximum distance between the light spot center and the droplet edge, as a function of the light intensity (e) and of the substrate temperature, for I = 100 W/cm$^2$ (f).

**Figure 3:** Video frames at different instants showing a fluid jet following the motion of the light spot**.** Jet's walking occurs through a continuous emission of new streams of fluid from its tip toward the illuminated area. T = 120°C, I = 100 W/cm$^2$.

**Figure 4:** Effect of **LN** focused irradiation close to a RM734 sessile droplet in the $N_F$ phase, on the DOWN side of the substrate. The beam positioned close to the droplet rim produces the emission of a single fluid jet in correspondence of the irradiated area (a). The jet spontaneously retracts back to the mother droplet during light irradiation (b) and (c). A small secondary droplet generated by the jet during its retraction is indicated by a blue arrow in (d). (e) Secondary droplets left on the LN surface after jet retraction in a different region repelled by the center of the illuminated area. T = 120°C, I = 100 W/cm$^2$.

**Figure 5:** Long fluid jet emitted by a sessile RM734 $N_F$ droplet upon focused illumination of an area within the droplet perimeter, on the DOWN side (a). By shifting the light spot in correspondence of the droplet rim, the jet elongates (b) while its irradiated section becomes thinner and thinner (c) and (d), up to the separation from the original droplet (e). The motion of the "free" fluid jet can then be optically controlled by moving the light spot close to its tip (f). T = 120°C, I = 100 W/cm$^2$.